\documentclass[12pt,preprint]{aastex}

\usepackage{graphicx}
\usepackage{amsmath} 
\usepackage{amsmath}

\shorttitle{Dynamical friction in the dwarf spheroidal in Fornax}
\shortauthors{Cowsik \& al.}

\begin{document}

\title{Internal dynamics and dynamical friction effects in the dwarf spheroidal galaxy in Fornax}

\author{Ramanath Cowsik\altaffilmark{1}, Kasey Wagoner\altaffilmark{1}, Emanuele Berti\altaffilmark{2,}\altaffilmark{3} and Amit Sircar\altaffilmark{1}}

\altaffiltext{1}{McDonnell Center for the Space Sciences, Department of
  Physics, Washington University, St. Louis, Missouri 63130, USA}
\altaffiltext{2}{Department of Physics and Astronomy,     
The University of Mississippi, University, MS 38677-1848, USA}
\altaffiltext{3}{Jet Propulsion Laboratory, California Institute of
  Technology, Pasadena, CA 91109, USA}

\begin{abstract}

  In the Fornax dwarf spheroidal galaxy the globular clusters are distributed
  widely, without any significant central concentration.  Oh et al. (2000)
  pointed out that such a distribution is paradoxical: dynamical friction
  effects estimated using single-component King models would have forced the
  globular clusters to spiral down to the center of the galaxy well within a
  Hubble time.  This paper is devoted to a discussion of this paradox.  We
  describe a model in which the stars of the dwarf spheroidal galaxy are
  embedded in a cloud of dark matter, and each of these components is
  specified by its own phase space distribution function.  This model allows
  us to fit self-consistently the observed luminosity profile and the spatial
  variation of the velocity dispersion of the stars.  This fitting yields two
  basic parameters, related to the central density and velocity dispersion,
  that characterize the phase space distribution of dark matter.  The
  dynamical friction effects calculated on the basis of this self-consistent
  model are small enough that the observed spatial distribution of the
  globular clusters poses no difficulty, and the apparent paradox is resolved.
  Thus we have at hand a model for Fornax that reproduces the main observed
  features of this dwarf spheroidal galaxy.

\end{abstract}



\section{Introduction}
The masses and the phase space structure of the smallest stellar systems that
form in the Universe provide the most direct clues to decipher the nature of
dark matter and the process of galaxy formation.  Among these systems, dwarf
spheroidal galaxies are the most suitable for the study of dark matter.  They
have low densities of visible matter in the form of stars (about a million
times smaller than those encountered in globular clusters of comparable
masses), and both their internal and external dynamics are dominated by the
cloud of dark matter (more commonly called the halo) in which they are
embedded.  Their stability against tidal disruption and also their internal
dynamics depend upon details of the structure and extent of the dark-matter
cloud, which in turn is sensitive to the phase-space structure of the dark
matter particles constituting the halo.

In recent years, improved astronomical observations of these faint systems
have become available for comparison with theoretical studies. One of the
earliest studies of these systems is by \cite{faber}, who used luminosity
profile observations to show that dwarf galaxies are dominated by dark matter
even in their cores, drawing attention to dwarf spheroidal galaxies as
candidates for the study of dark matter. Around the same time, \cite{aaronson}
measured the radial velocities for several carbon stars in these systems with
sufficient accuracy to support the idea that dark matter plays a dominant role
in their internal kinematics.  Prompted by these studies,
\cite{cowsik1,cowsik2} developed an ``embedding model'' where the dwarf
spheroidals are embedded in an extensive cloud of dark matter.  A similar
approach was championed by \cite{pryor} in the context of dwarf spheroidal
galaxies.  \cite{tremaine} and \cite{hernandez} discussed dynamical friction
effects on globular clusters in dwarf galaxies. They derived simplified
analytical formulae for the orbital evolution of a massive body in a dark halo
which is assumed to follow the King distribution \citep{king,binney}, finding
that the dynamical timescales for the globular clusters to sink to the center
of the galaxy are about 1~Gyr, and that these timescales increase as the core
radius increases.

Of particular relevance to the present study is the work of \cite{oh}, who
pointed out that the five globular clusters gravitationally bound to the dwarf
spheroidal galaxy in Fornax paradoxically ``preserve their diffuse spatial
distribution despite the fact that the clusters' orbital decay timescale is
much shorter than the estimated age of the host galaxy.''  \cite{sanchez} have
shown that the problem is exacerbated if modified Newtonian dynamics is
applicable.  There is a similar problem with the orbital decay timescales in
dwarf elliptical galaxies \citep[for details see e.g.][]{lotz}. \cite{goerdt}
considered the effects of dynamical friction in dark matter halos with
profiles having a central cusp, as suggested by cosmological simulations
\citep{hernquist,navarro,moore} and showed that ``in a cuspy CDM halo the
globulars would sink to the center from their current position within a few
Gyrs, presenting a puzzle.'' They further point out that should the stellar
population be embedded in a dark matter halo with a large core, this
difficulty would be resolved.  In making this point, \cite{goerdt} assume that
the dark matter halo has a density profile given by
\begin{equation}\label{goerdt_density}
\rho(r)=\rho_D(0)[1+(r/r_s)]^{-3}
\end{equation}
\noindent with a central density $\rho_D(0)=0.1{\rm M_\odot pc^{-3}}$ and
scale size $r_s=2.4{\rm kpc}$.  They also assume that the density sharply cuts
off at $\sim 50{\rm kpc}$.  Thereafter Jeans' equations representing the
balance between gravitational forces and the gradient in pressure provides
them with the $\langle v^2(r)\rangle$ for this assumed density profile. This
allows them to estimate the time scales for the dynamical friction to operate
and conclude that in such a halo the dynamical friction effects on the
globular clusters are small.  This immediately opens up the following
questions: (i) Does the density distribution for dark matter given in
Eq.~\ref{goerdt_density}, with the chosen values for the parameters
$\rho_D(0)$ and $r_s$, represent correctly the density profile of the dark
matter halo in Fornax?  (ii) Would the stars embedded in such a halo follow
the observed spatial distribution?  (iii) Would the observed $v_{\ast rms}(r)$
of these stars be correctly reproduced?

In other words, in order to explicitly resolve the problem posed by the wide
distribution of globular clusters in Fornax, we should work within the context
of a model that will allow us to formally derive the density profile
$\rho_\ast(r)$ of the stars and the profile of their velocity dispersion
$v_{\ast rms}(r)$ in terms of the parameters describing the phase space
distribution of dark matter.  Such a phase space distribution, with the best
choice for the parameters, may then be inserted into the relevant integral in
Chandrasekhar's formula to estimate the time constant for the migration of
globular clusters toward the center of Fornax.  As a follow up of the earlier
work of \cite{tremaine}, \cite{hernandez}, and \cite{goerdt}, this paper
describes an attempt to construct such a model and address the problem posed
by the wide distribution of globular clusters in Fornax.

\begin{figure}
\begin{center}
 \includegraphics*[width=0.5\textwidth]{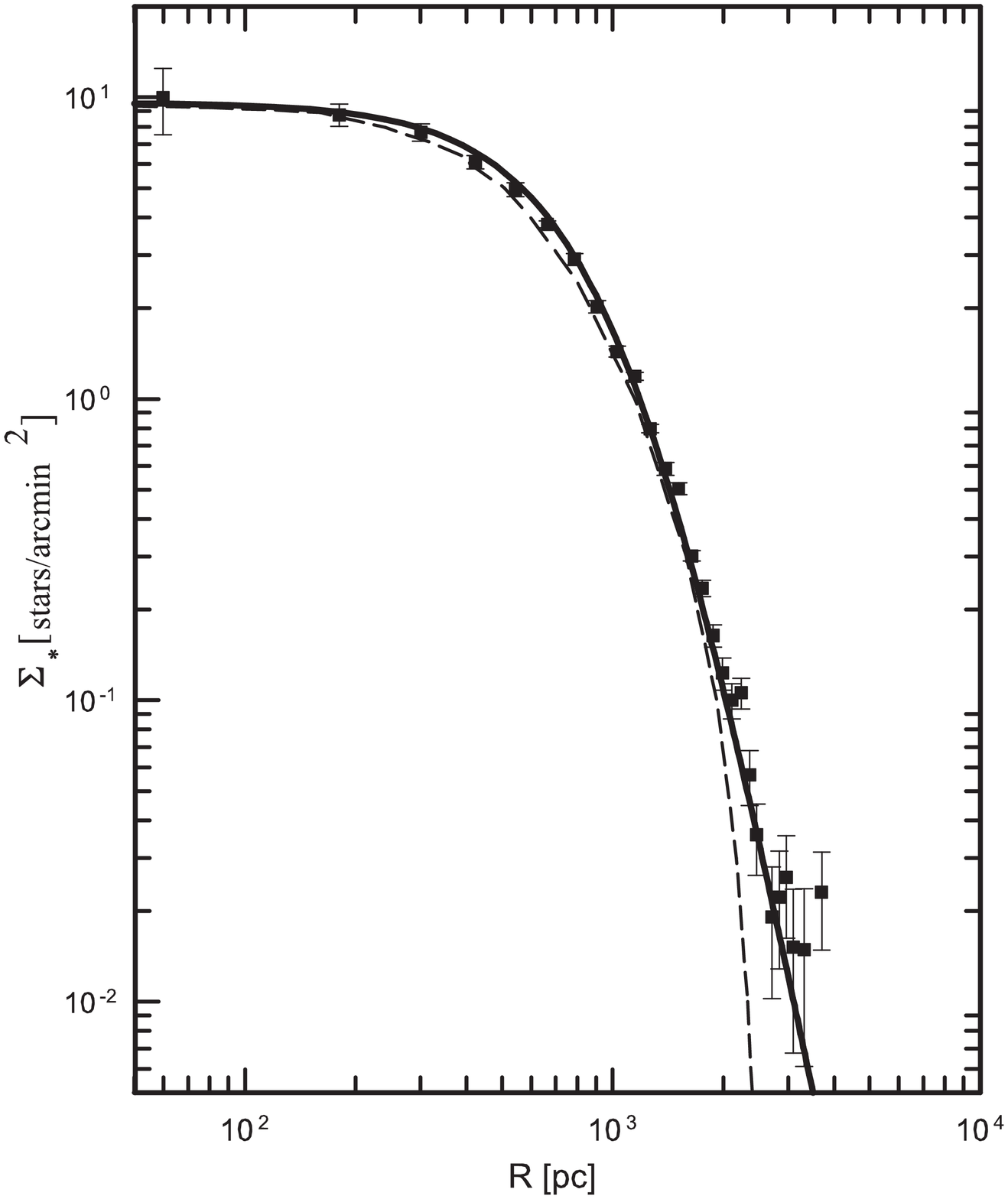} \figcaption[]
                  {\label{fig:coleman} The theoretically calculated surface
                    density in Eq.~\ref{rhoast} (solid black line) is compared
                    with the observations by \cite{coleman} and their single
                    component King model fit (dashed line corresponding to
                    $\sigma_\ast \approx 11\rm{km\cdot s^{-1}}$ and
                    $r_{k\ast}\approx 2.6$kpc), which progressively falls
                    below the observed surface densities at distances beyond
                    the core radius. The embedding model (solid black line)
                    provides a very good fit.}
\end{center}
\end{figure}

\begin{figure}
\begin{center}
  \includegraphics*[width=0.5\textwidth]{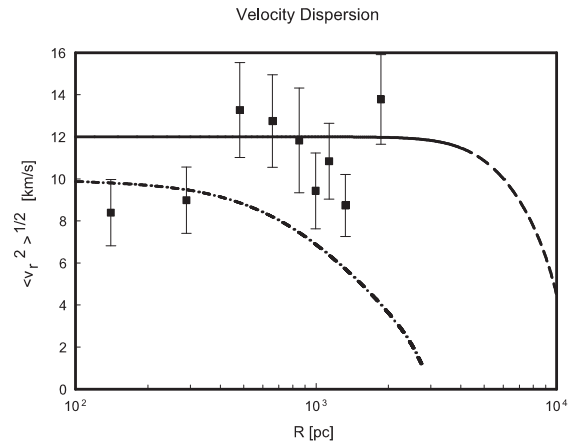} \figcaption[]
                   {\label{fig:walker} The observed dispersion in the radial
                     velocities using 182 stars \citep{walker} is compared
                     with our theoretical model (solid black line).  A single
                     component King model presented by Walker et al. (dash-dot
                     line) with $\phi(0)/\sigma_\ast^2=3.26$ and tidal radius
                     $r_{k\ast}\approx 3$kpc predicts smaller dispersion
                     velocities beyond the core radius.}
\end{center}
\end{figure}

Fortunately aiding the model building effort, astronomical observations of the
dwarf spheroidal galaxy in Fornax have improved considerably since the
pioneering studies by \cite{hodge1,hodge2}, and now excellent data sets are
available. The galaxy is located at a distance of $\sim$138~kpc
\citep{mateo}. It covers a large area of the sky, with an estimated tidal
radius of $\sim$71' \citep{irwin}, corresponding to $\sim 2.85$~kpc.  A
careful wide-field survey by \cite{coleman} provided a precise radial number
density profile of the RGB-selected stars, showing that the observed
distribution increasingly deviates from a single-component King distribution
beyond the core radius (see Fig.~\ref{fig:coleman}).  The surface density
beyond the core radius falls off as a power law and significant densities were
noted up to ~90', far beyond the earlier estimates of the tidal
radius. \cite{walker} measured the radial velocities of a large number of
stars in the galaxy and derived the radial velocity dispersion $\langle
v_r^2(R)\rangle^{1/2}$ as a function of the projected distance $R$.  They
noted that the radial dependence is flat, again deviating significantly from
the decreasing behavior as a function of $R$ that would be expected for a
single-component King distribution (Fig.~\ref{fig:walker}).  The five globular
clusters that pose a problem for dynamical friction timescales are located at
projected distances $R=0.24,~0.43,~1.05,~1.43$ and $1.60$~kpc, respectively
\citep{walker}. Of course their actual radial distances would be larger, by a
factor of $\sim \sqrt{3/2}$ on the average.

In order to model the observations noted above, we derive in Section
\ref{model} the radial dependence of the velocity dispersion and the surface
density profile of stars within the context of a dynamical model, by
formulating the Boltzmann-Poisson equations in the embedding approximation.
The solution to these equations is obtained through a standard procedure,
which is briefly described. We assume that the underlying phase space
distributions, both for the dark matter and for the stars, are nearly
isothermal and follow the lowered isothermal distributions \citep{king}. In
making this choice we were guided by extensive earlier work adopting lowered
isothermal phase space distributions to model these systems (see for example
\cite{irwin}, \cite{binney} and references therein), as well as by the
requirement of simplicity.  Since there are two components in the Fornax
system - the dark matter particles and the stars - we need at least two
distributions to specify them, and the reduced isothermal distributions have
the advantage that they form cored systems with finite masses and have been
adopted extensively before for modeling various astronomical systems,
including the dwarf spheroidal galaxies.  It is perhaps relevant to point out
here that recently some concerns have been expressed regarding their
applicability to galactic systems \citep{gilmore}.  These authors have noted
that the two-body relaxation time in these systems, $\tau_2\approx (8N/\log
N)\tau_{cross}$, is much longer than the age of the universe, and as such the
system would not have relaxed to a nearly Maxwellian phase space distribution.
This remark is true and the absence of any segregation of massive stars into
the central regions (except in the case of the densest globular clusters)
indeed points to the absence of such equilibrium induced by two-body
interactions.  However, the smooth density profiles of these systems and our
ability to model them with reduced isothermal distribution functions beg for
an explanation. The pioneering study by \cite{lynden} has shown that the
equilibria in such systems are dictated by collective effects.  He showed that
even though the `fine-grained' distribution takes almost infinite time to
evolve, the `coarse-grained' distribution function rapidly relaxes due to
collective effects into a quasi-stationary state - a process called `violent
relaxation'.  His work triggered wide-ranging studies of the collective
effects of long-range forces.  As long as initial conditions avoid parametric
oscillations and kinetic energy approximately balances the potential energy,
$2T+U\approx 0$, the coarse grained distribution does turn out to be
quasi-Maxwellian \citep{levin}.  It may be noted further that it is the
fluctuation of the gravitational potential due to collective effects that
brings about this quasistationary state: all stars acquire the same
distribution of velocities and there is no segregation, in contrast to
equilibria driven by two body interactions.  Accordingly we adopt reduced
isothermal distribution functions for the stars and dark matter particles,
solve self-consistently the equations of the embedding model and therefrom
compute the slowdown rate due to dynamical friction.  In Section
\ref{observations} we compare the theoretical predictions with the observed
velocity dispersion and luminosity profile, and we show that the model indeed
fits the observations very well.  Thus the main parameters of our model (a
density parameter $\rho_{CD}$ and a velocity dispersion parameter
$\sigma_{D}$) are determined with sufficient accuracy, allowing us to estimate
with confidence the time needed for the globular clusters to sink to the
center due to dynamical friction.  This turns out to be longer than the Hubble
time, thereby resolving the paradox. Section \ref{conclusions} is devoted to a
general discussion and concluding remarks.

\section{The Dynamical Model}
\label{model}

We begin the theoretical analysis by developing a dynamical model for the
dwarf galaxy in Fornax.  A comparison of the predictions of the model with the
observations of the profiles of the root mean square of the line of sight
velocities $v_{rms\ast}(R)$, and the projected number density,
$\Sigma_\ast(R)$, of the stars shows that the model provides an excellent fit
to the observations and provides the best values of the two basic parameters
$\rho_{CD}$ and $\sigma_{D}$ characterizing the phase space distribution
function of the dark matter particles in Fornax.  Using these we analytically
integrate the incomplete integral over the phase space distribution appearing
in the Chandrasekhar formula for dynamical friction.

Let the phase space distribution both of the dark matter $f_D$ and of the
stars $f_\ast$ be of the `reduced isothermal' form (King 1966)
\begin{eqnarray}\label{reduced_iso}
f_i&=&\rho_{Ci}(2\pi \sigma_i^2)^{-3/2}\left[e^{\varepsilon_i/\sigma_i^2}-1\right] \hspace{3mm} {\rm for}\hspace{2mm} \varepsilon_i>0\,,\nonumber\\
&=&0\hspace{3mm} {\rm for}\hspace{2mm}\varepsilon_i\leq 0\,.
\end{eqnarray}
\noindent The subscript $i$ could either be `$D$' or `$\ast$' for the
particles of dark matter or the stars as appropriate, and the `binding energy'
$\varepsilon$ is given by
\begin{equation}\label{binding_energy}
\varepsilon_i=\left\{\phi(r_{Ki})-\phi(r)-v^2/2\right\}\equiv\left\{\xi_i-v^2/2\right\}\,,
\end{equation}
\noindent where $\phi(r)=\phi_D(r)+\phi_\ast(r)$ is the total gravitational
potential deriving contributions from dark matter and the stars.  The terminal
radii $r_{Ki}$ for the two components are indirectly specified by the choice
of the value of the total potential $\phi$ at these locations. Since the total
gravitational potential due both to the dark matter and the stars occurs in
both the distribution functions, it couples the two components.  Notice that
these distribution functions depend only on $v^2$ which implies isotropy for
the velocity distribution, and the distribution functions, being functions of
energy alone, satisfy the stationary collisionless Boltzmann equation.  The
parameter $r_{Ki}$ represents the apogalacticon at which the density vanishes,
and this is indirectly specified by the choice of $[\phi(r_{Ki})-\phi(0)]$.
The integral over the velocities \citep{king,binney} provides the spatial
density $\rho_D(r)$ and $\rho_\ast(r)$:
\begin{equation}\label{spatial_density}
\rho_i=\rho_{Ci}\left[e^{y_i}{\rm erf}(y_i^{1/2}) - \sqrt{\frac{4y_i}{\pi}}(1+\frac{2}{3}y_i)\right]\,,
\end{equation} 
\noindent where $y_i=\xi_i/\sigma_i^2$ are the scaled potentials satisfying
the spherically symmetric Poisson equations
\begin{equation}\label{poisson}
\frac{2}{r}\frac{dy_i}{dr}+\frac{d^2y_i}{dr^2}=-\frac{4\pi G}{\sigma_i^2}\rho_i(r)\,.
\end{equation}
\noindent Notice that both $\rho_D(r)$ and $\rho_\ast(r)$ are functions of the
total potential $\phi(r)=\phi_D(r)+\phi_\ast(r)$.  This represents the
coupling between the two systems; as the density of one of these becomes
higher its contribution increases and the total potential becomes deeper,
making both components more compact.  Accordingly, these equations represent
two coupled non-linear differential equations, which are solved by Taylor
expansion of $y_i$ about the origin, say up to the sixth order in $r$, and
thence integrated using standard numerical methods (an adaptive Runge-Kutta
scheme) to extend the solutions from small values of $r$ to any desired
distance.  Noting that the mass to light ratio of dwarf spheroidals, including
that in Fornax, are very large, we are able to simplify the numerical analysis
by neglecting the contribution of the stars to the overall gravitational
potential $\phi(r)$.  From the numerical values of $y_D$ and $dy_D/dr$ all
quantities of interest may be evaluated: for example the circular velocity
$v_c(r)=\{\sigma_D^2r|dy_D/dr|\}^{1/2}$, and the mass of dark matter contained
within a radius $M_D(r)=\sigma_D^2r^2|dy_D/dr|/G$.  Furthermore, noting that
with dominance of dark matter
\begin{equation}\label{scaled_potential}
y_\ast(r)=\left[y_D(r)-y_D(r_{K\ast})\right]\sigma_D^2/\sigma_\ast^2\,,
\end{equation}
\noindent the density of the stars is evaluated using
Eq.~\ref{spatial_density}.  Now, the projection of any variable $Z(r)$ on the
plane of the sky is given by
\begin{equation}\label{S}
S(Z,R)=\int_{R}^{r_{max}} 2Z(r)\frac{rdr}{(r^2-R^2)^{1/2}}\,,
\end{equation}
\noindent where $R$ is the projected radial distance.  With this, the surface
density profile of stars is given by
\begin{equation}\label{rhoast}
\Sigma_\ast(R)=S(\rho_\ast(r),R)\,.
\end{equation}
\noindent Defining the second moment of the phase space distribution of stars
as
\begin{eqnarray}\label{Tast}
T_\ast(r)&=&\int_0^{\sqrt{2\xi_\ast}}f_\ast v^2\cdot 4\pi v^2 dv\nonumber\\
&=&3\rho_{C\ast}\sigma_\ast^2\left[e^{y_\ast}{\rm erf}(y_\ast^{1/2})-\frac{8}{3\sqrt{\pi}}\left\{\frac{y_\ast^{5/2}}{5}+\frac{1}{2}y_\ast^{3/2}+\frac{3}{4}y_\ast^{1/2}\right\}\right]
\end{eqnarray}
\noindent and recalling the assumption of isotropy,
\begin{equation}\label{vast}
\left\langle v_\ast^2(r)\right\rangle=\frac{T_\ast(r)}{3\rho_\ast(r)}\hspace{2mm} {\rm and}\hspace{2mm} v_{rms,\ast}(R)=\left[\frac{1}{3}\frac{S(T_\ast,R)}{S(\rho_\ast,R)}\right]^{1/2}\,.
\end{equation}

In Eqns.~\ref{S}-\ref{vast} we have the theoretical functional forms for the
profile of stellar surface density $\Sigma_\ast(R)$ and the profile of stellar
velocity dispersion $v_{rms \ast}(R)$, which may be compared with the
corresponding profiles obtained from observations to determine the best choice
of the parameters entering the distribution functions.

Now, Chandrasekhar's formula \citep{chandra,binney} provides an estimate of
the dynamical friction acting on the globular clusters in Fornax:
\begin{equation}\label{chandradf}
\frac{d\vec{V}}{dt}=-\frac{\vec{V}}{|V|^3}4\pi G^2 M\ln \Lambda 
\int_0^{V_m}f_{_D}4\pi v^2dv\,,
\end{equation}
where we have set $(M+m)\approx M$, and estimate the Coulomb logarithm as
$\ln\Lambda=4$; also here $V_m$ is the smaller of $V$ and $[2\xi(r)]^{1/2}$,
the latter being the escape velocity from the system.
Defining $x_m^2=V_m^2/2\sigma_{_D}^2$ we get
\begin{eqnarray}\label{dynfri}
\frac{d\vec{V}}{dt}&=&
-\frac{\vec{V}}{V^3}4\pi G^2M \rho_{_D}\ln\Lambda\, \\
&\times& \left[e^{y_D}\left\{{\rm erf}(x_m)-
\frac{2x_m}{\sqrt{\pi}}e^{-x_m^2}\right\}-\frac{4}{3\sqrt{\pi}}x_m^3\right]\,.
\nonumber
\end{eqnarray}

Thus, we have at hand the formalism needed to fit the observed profiles of
number counts and of dispersion in the stellar velocity and obtain the
parameters $\rho_D(0)$ and $\sigma_D$ crucial to the phase space distribution
function of the dark matter particles.  These will allow us to use
Eq.~\ref{dynfri} to evaluate the dynamical friction effects on the globular
clusters in Fornax.

\section{Comparison with observations}
\label{observations}


The theoretical profiles $v_{rms \ast}(R)$ and $\Sigma_\ast(R)$ depend on the
choice of the model parameters: $\rho_{CD}$ (or more conveniently
$\rho_D(0)$), $\sigma_D$, $r_{KD}$ for the dark matter and $\rho_{C\ast}$ (or
$\rho_\ast(0)$), $\sigma_\ast$ and $r_{K\ast}$ for the stars.  Amongst these
$\rho_\ast(0)$ provides the normalization constant for $\Sigma_\ast(R)$ for
the particular type of stars which are being counted, and $r_{K\ast}$ the
location where we would like to cut-off the stellar distribution.  Noting that
the farthest data point in the stellar profile observed by \cite{coleman} lies
at a projected distance $R\approx4{\rm kpc}$, we may choose $r_{K\ast}$ to be
any value significantly greater than this, say $10{\rm kpc}$.  The observed
profile is insensitive to this choice.  Similarly the dependence of $r_{KD}$
is also weak, fixing the total mass of the dark matter halo.  Thus the crucial
parameters that control the theoretical predictions are $\sigma_\ast$,
$\sigma_D$ and $\rho_D(0)$.  It may be appropriate to point out here that
$y_D(0)$ and $y_\ast(0)$ are not additional independent parameters.  The
choice of $\rho_D(0)$, $\sigma_D$, $r_{KD}$ and corresponding stellar
parameters fixes them through the requirement that $y_D(r_{KD})=0$ and
$y_\ast(r_{K\ast})=0$.


We start by fitting the observed luminosity profile $\Sigma_\ast(R)$ based on
the embedding model presented in the earlier sections.  Our fit is shown in
Fig.~\ref{fig:coleman} for the following set of parameters:
\begin{equation}{\label{parameters}}
\rho_D(0)=0.04{\rm M_\odot pc^{-3}}, \hspace{5mm} \sigma_D=19.5{\rm km\cdot s^{-1}}, \hspace{5mm}\sigma_\ast=12{\rm km\cdot s^{-1}}\,.
\end{equation}
\noindent Note that the model fits the observations very well, including the
power law tail at large radii.  In contrast, a single component King model
does not fit the observations, as noted by \cite{coleman}.  We emphasize that
the fit to $\Sigma_\ast(R)$ is very sensitive to the choice of $\rho_D(0)$,
$\sigma_D$, and $\sigma_\ast$, and at most a ten percent variation may be
accommodated in the values of these parameters given in Eq.~\ref{parameters}.
The values of $\rho_{_D}(0)$ and $\sigma_\ast$ determine approximately the
core radius of the stellar distribution in the embedding model; this radius,
$r_{0\ast}=[9\sigma_\ast^2/4\pi G\rho_D(0)]^{1/2}=650{\rm pc}$, marks the
location where the projected stellar density falls to about one half of its
central value. It is worth noting here that the quantity
$(2\sigma_{_D}^2/\sigma_\ast^2)$ yields approximately
$d\ln\rho_\ast(r)/d\ln(r)$, i.e. the power law exponent of the fall off of
$\rho_\ast(r)$ beyond the core; or equivalently,
$[(2\sigma_{_D}^2/\sigma_\ast^2)-1]$ gives the index for $\Sigma_\ast(R)$.

In order to ensure that our model is consistent with all the rest of the
available observations we calculate $\langle v_{r\ast}^2(R)\rangle^{1/2}$ with
the same set of parameters fixed by fitting $\Sigma_\ast(R)$, and compare the
prediction with the observed radial velocity dispersion \citep{walker}. The
data and our fit are shown in Fig.~\ref{fig:walker}.  Even though the
statistical uncertainties in the data set are large, our model is consistent
with the observations.  On the other hand, as pointed out by Walker et al.,
their observations are inconsistent with the single component King model
(dashed line).

Having fixed the parameters of the model, it is a straightforward matter to
calculate dynamical friction effects on the globular clusters using
Eq.~\ref{dynfri}. A convenient way to display the results is to assume
circular orbits at various radial distances of closest approach and
calculate the dynamical friction rate for $V=v_c$, where $v_c$ is the circular
velocity.  There are two reasons for doing so: (1) if dynamical friction
effects were indeed significant, then the orbits of the globular clusters
would tend to be circularized by friction; (2) dynamical friction effects are
maximized in the circular case, because non-circular orbits with the same
distance of closest approach have higher velocities, and they also sample the
outer, low-density regions of the dark matter distribution.

\begin{figure}
\begin{center}
  \includegraphics*[width=0.5\textwidth]{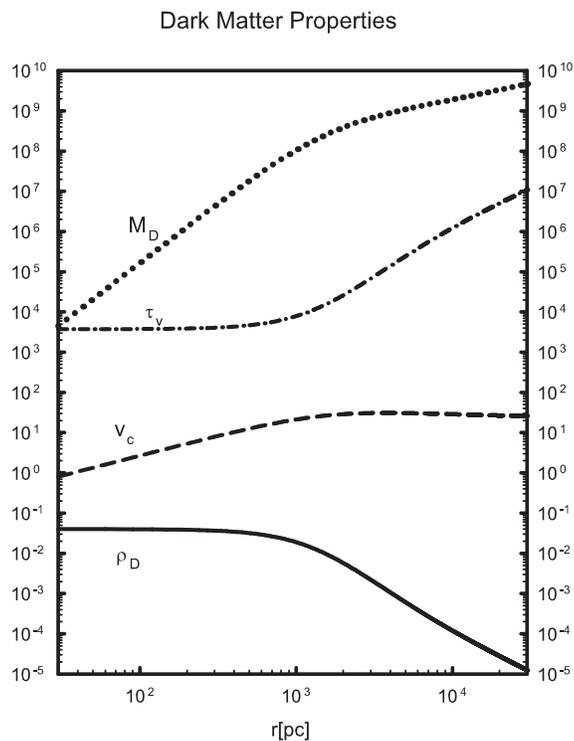} \figcaption[]
                   {\label{fig:dm_properties} The lifetime
                     $\tau_v=|V/(dV/dt)|$ for orbital decay due to dynamical
                     friction, in Myr (dash-dotted line) is calculated for
                     circular orbits, using the best-fit parameters for the
                     dark matter distribution obtained from the observed
                     stellar velocity dispersion and stellar density
                     profiles. For completeness we also show the enclosed mass
                     in dark matter $M_{_D}$ in units of ${\rm M}_\odot$
                     (dotted line), the circular speed $v_c$ in ${\rm km}/{\rm
                       s}$ (dashed line) and the density of dark matter
                     $\rho_{_D}$ in ${\rm M_\odot}/{\rm pc}^3$ (solid line).}
\end{center}
\end{figure}

Choosing the best-fit parameters, $\rho_D(0)=0.04{\rm M_\odot pc^{-3}}$ and
$\sigma_D=19.5 {\rm km\cdot s^{-1}}$, we calculate the rate of dynamical
friction given in Eq.~\ref{dynfri} and plot it in
Fig.~\ref{fig:dm_properties}. Very close to the center (i.e., for small $r$)
the dynamical friction formula (\ref{dynfri}) simplifies to
\begin{equation}\label{force}
\frac{d\vec{V}}{dt}\approx
-\frac{16\pi^2 G^2 M \ln\Lambda \rho_{_D}(0)}{3(2\pi\sigma_{_D}^2)^{3/2}}\vec{V}\,,
\end{equation}
which implies a lifetime
\begin{equation}\label{lifetime}
\tau \approx
\left|\frac{V}{dV/dt}\right|=
\frac{3\sigma_{_D}^3}{4\sqrt{2\pi}G^2  M \ln \Lambda \rho_{_D}(0)}
\approx 4.5~{\rm Gyr}
\end{equation}
for globular clusters in Fornax.  
This simple estimate is consistent with the exact calculations presented in
Fig.~\ref{fig:dm_properties}. The effective lifetime, $\tau$, is $\sim
4.5$~Gyr in the central parts of the dwarf spheroidal galaxy, it is nearly
constant out to the core radius, and for larger radii it increases roughly
quadratically with $R$.  Since it would take several $e$-folding times for a
globular cluster to sink to the center, even a globular cluster initially
formed in the core region would not have migrated to the center within a
Hubble time.  For globular clusters that condensed in the outer regions or
with elongated orbits, the effects of dynamical friction are much weaker than
this estimate.

To be more specific, the dynamical friction causes loss of angular momentum at
a rate
\begin{equation}
\frac{dL}{dt}=r\frac{dv}{dt}\,.
\label{eq:ang_momentum}
\end{equation}
\noindent Here the angular momentum $L$ is taken to be $r\cdot v_c$, and the
dynamical friction $dv/dt$ is evaluated using Eq.~\ref{dynfri} with $V=v_c$,
keeping in mind that friction will tend to make the orbits circular.  For a
given perigalacticon circular orbits suffer the highest rate of dynamical
friction, as elliptical orbits will spend a large part of their orbital time
in the lower density outer regions.  In order to calculate the radial
migration rate, we write
\begin{equation}
\frac{dr}{dt}=\biggl(\frac{dL}{dt}\biggl)/\biggl(\frac{dL}{dr}\biggl)=\biggl(r\frac{dv}{dt}\biggl)/\frac{d}{dr}(r\cdot v_c)
\label{eq:rdot}
\end{equation}
\noindent and define the radial migration time constant $\tau_r$ by 
\begin{equation}
\tau_r(r)\approx r/\left|\frac{dr}{dt}\right|\,.
\label{eq:time_constant}  
\end{equation}

\begin{figure}
\begin{center}
  \includegraphics*[width=0.5\textwidth]{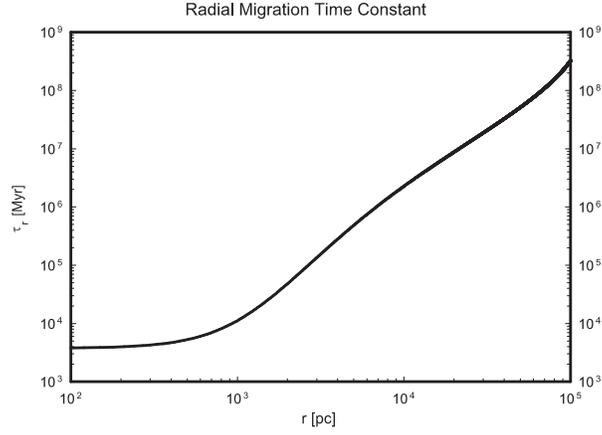} \figcaption[]
                   {\label{fig:migration_times} The radial migration time
                     constant $\tau_r=r/|dv/dt|$ is displayed as a function of
                     $r$.}
\end{center}
\end{figure}

\begin{figure}
\begin{center}
  \includegraphics*[width=0.5\textwidth]{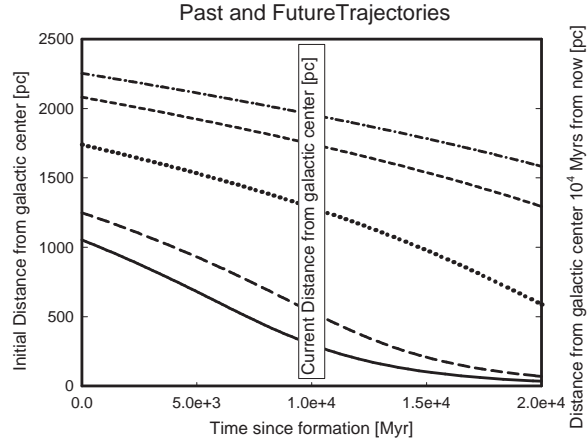} \figcaption[]
                   {\label{fig:trajectories} The radial trajectories of
                     globular clusters from the time of the formation up to
                     10Gyr in the future is shown.  The trajectories pass
                     through their present location, $r=\sqrt{3/2}R$, 10Gyr
                     after formation.}
\end{center}
\end{figure}

This quantity is plotted in in Fig.~\ref{fig:migration_times}.  Note that
$\tau_r$ becomes very large in the outer region, and becomes equal to $\tau_v$
well within the core radius.  Indeed, the radial migration rate given in
Eq.~\ref{eq:rdot} is easily integrated to obtain $r(t)$ for various assumed
values of $r_n$, the present location of the globular clusters in Fornax.  We
show in Fig.~\ref{fig:trajectories} the past and the future trajectories,
starting from the time of their formation $\sim 10^4$ million years ago,
arriving at their current location in the present epoch, and moving into the
future another $10^4$ million years.  These trajectories exclude any
paradoxical behavior and are consistent with the astrophysical assessment of
the regions of their formation and subsequent evolution of their orbits.  They
show that the globular clusters that were formed at distances between $\sim 1$
and 2 kpc from the center of Fornax, suffered dynamical friction at the
expected rate and arrived at their current locations at the present epoch.
During the next 10 billion years they will migrate closer to the center, but
will remain distinct without forming a nucleus.

\begin{figure}
\begin{center}
  \includegraphics*[width=0.5\textwidth]{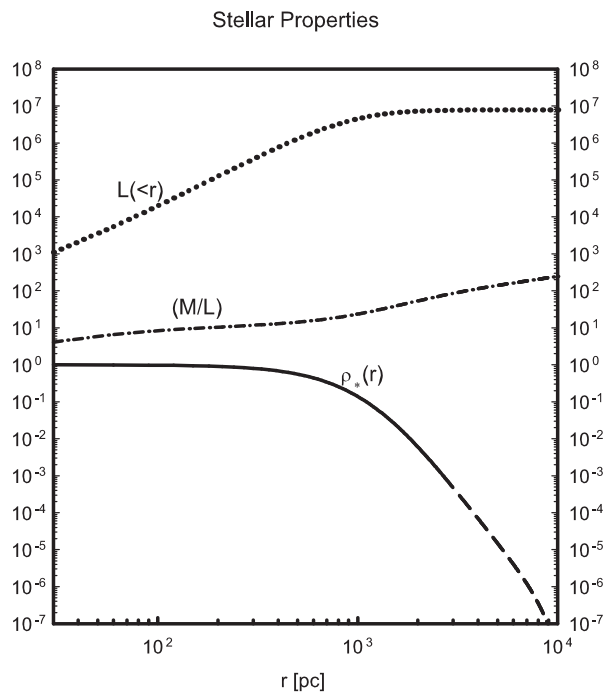} \figcaption[]
                   {\label{fig:stellar_properties} The theoretical estimates
                     of the stellar density $\rho_\ast(r)$ in ${\rm
                       M}_\odot$/pc$^{3}$ (solid line), stellar luminosity $L$
                     enclosed within a radius $r$ in ${\rm L}_\odot$ (dotted
                     line) and the total mass to luminosity ratio $M/L$ also
                     enclosed within $r$ in units of ${\rm M}_\odot/{\rm
                       L}_\odot$ (dash-dotted line) are shown as a function of
                     the radial coordinate.  Note that the density of stars
                     becomes less than one thousandth of its central value for
                     $r>3$kpc, and it is unclear whether these stars really
                     belong to the dwarf galaxy.  This is indicated by the
                     dashed-line extension to the model estimates of
                     $\rho_\ast(r)$.}
\end{center}
\end{figure}

Thus the apparent paradox posed by the diffuse distribution of globular
clusters in the Fornax dwarf spheroidal galaxy \citep{oh} is resolved.  In
order to provide some of the derived parameters of the stellar component in
Fornax, we take the total visible magnitude of Fornax $m_L\approx9.3$
\citep{karachentsev}, distance $d=138$kpc and a bolometric correction $\sim
1^m$ to estimate the total luminosity of $L\approx 7.5\times 10^6 L_\odot$.
(For a bolometric correction of $\sim 0.5^m$, $L\approx 6\times 10^6
L_\odot$).  In Fig.~\ref{fig:stellar_properties} we show the density of the
stars $\rho_\ast(r)$ in arbitrary units, the luminosity within a radius $r$
and the mass to luminosity ratio as a function of the radius; of course the
mass is dominated by the dark matter.

\section{Discussion and conclusions}
\label{conclusions}

The dynamical friction suffered by astronomical objects (such as globular
clusters) depends not only on the local spatial density of the surrounding
region, but also on its phase space structure.  Since the phase-space density
is invariant along a dynamical trajectory, this sensitivity implies that
several parameters of interest (such as the dispersion velocities and the
extent of the dark matter halo) may be determined by examining the role played
by dynamical friction in extended structures in the Universe.  These remarks
are illustrated here through a study of the dynamics of globular clusters
surrounding the dwarf spheroidal galaxy in Fornax. To summarize, a model based
on reduced isothermal distributions both for the stars and the dark matter
provides a very good fit to the radial velocity dispersions $v_{rms\ast}(R)$
and the luminosity profile $\Sigma_\ast(R)$ observed for the stars in the
dwarf spheroidal in Fornax.  These fits determine $\rho_D(0)$ and $\sigma_D$,
the two crucial parameters of the phase space distribution of dark matter.
The rates of dynamical friction calculated using these parameters are
sufficiently small that the paradox pointed out by \cite{oh} is resolved.  An
application of such models to other systems including galaxy clusters would
provide further insight into the phase space distribution of dark matter and
the process of structure formation in the universe.  It would also be useful
to attempt self-consistent fits to the observations of Fornax, using the phase
space distributions obtained with numerical simulations, as soon as they
become available.

\noindent Acknowledgments: We thank Dr. Tsitsi Madziwa-Nussinov for help with
the preparation of the manuscript, and Professor Francesc Ferrer for
discussions.


\newpage

\end{document}